\documentclass[usenatbib]{mnras}
\usepackage{graphicx}

\begin{document}

\title[Overconfidence in Photometric Redshifts]{Overconfidence in
  Photometric Redshift Estimation}

\author[David Wittman et al.]{David Wittman,$^{1,2}$ Ramya Bhaskar,$^{1}$ and Ryan Tobin$^{1,3}$ \\
$^{1}$Physics Department, University of California, Davis, CA 95616\\
$^{2}$Instituto de Astrof\'{i}sica e Ci\^{e}ncias do Espa\c{c}o,
Faculdade de Ci\^{e}ncias, Universidade de Lisboa, Lisbon, Portugal\\
$^{3}$Current address: Department of Physics and Astronomy,
  University of Hawaii, Honolulu, HI 96822}

\maketitle

\begin{abstract} 
  We describe a new test of photometric redshift performance given a
  spectroscopic redshift sample.  This test complements the
  traditional comparison of redshift {\it differences} by testing
  whether the probability density functions $p(z)$ have the correct
  {\it width}.  We test two photometric redshift codes, BPZ and EAZY,
  on each of two data sets and find that BPZ is consistently
  overconfident (the $p(z)$ are too narrow) while EAZY produces
  approximately the correct level of confidence.  We show that this is
  because EAZY models the uncertainty in its spectral energy
  distribution templates, and that post-hoc smoothing of the BPZ
  $p(z)$ provides a reasonable substitute for detailed modeling of
  template uncertainties.  Either remedy still leaves a small surplus
  of galaxies with spectroscopic redshift very far from the peaks.
  Thus, better modeling of low-probability tails will be needed for
  high-precision work such as dark energy constraints with the Large
  Synoptic Survey Telescope and other large surveys.
\end{abstract}


\section{Introduction}

Photometric redshifts are of key importance to current and future
galaxy surveys.  A variety of methods have been demonstrated, falling
broadly into two categories: empirical and template-based.  Empirical
methods predict redshifts from photometry by directly using the known
spectroscopic redshifts of a subsample spanning the color and
magnitude range of the main photometric sample.  Template methods use
models of galaxy spectral energy distributions (SEDs), which enable
prediction of redshifts beyond the magnitude limit of the
spectroscopic sample.  See \citet{HH10} and \citet{Dahlen2013zphot}
for overviews and performance comparisons.



Until recently, photometric redshift performance comparisons
\citep{Hogg98,HH08,HH10} have been based on casting the photometric
redshift of a galaxy as a single number, but this glosses over some of
the complexity inherent in these predictions.  For example, a deep
survey with a small number of filters is bound to encounter
degeneracies in which both low- and high-redshift models are
acceptable for some galaxies.  Forcing a photometric redshift
algorithm to choose only the single most likely model thus generates
some wildly inaccurate redshift estimates, which are called
``catastrophic outliers.''  Capturing all the photometric redshift
information in a probability density function $p(z)$ greatly reduces
or eliminates these outliers \citep{Fernandez02}.  Even if a
particular $p(z)$ is not multiply peaked, it may be asymmetric, so
that using the full $p(z)$ rather than a point estimate reduces bias
\citep{Mandelbaum08} and thus reduces systematic errors on downstream
science such as dark energy parameter estimation \citep{Wittman09}.

The works cited above established that the $p(z)$ paradigm offers
better performance than point estimates, but point estimates are still
more easily checked against spectroscopic redshift, by tabulating the
mean and scatter in the quantity $z_s-z_p$ (spectroscopic redshift
minus photometric redshift).  The $p(z)$ paradigm offers no obvious
generalization of this procedure.  Indeed, codes that work internally
with $p(z)$ often default to outputting a single error estimate for
each galaxy.  \citet{HH08} found that these error estimates are not
predictive of the real errors, but this may simply reflect the
underlying complexity of $p(z)$.  The extensive performance comparison
of \citet{Dahlen2013zphot} did use $p(z)$ to derive 68\% and 95\%
confidence intervals, and found that most codes are
overconfident---their confidence intervals are too narrow.  In this
paper we present a tool for systematically testing overconfidence, and
we show why the $p(z)$ output by template codes can be substantially
overconfident.  One limitation of our test is that spectroscopic
subsamples may not be representative of the full photometric sample.
However, this limitation also affects verification of point estimates,
and is therefore separable from the question of how to assess the
quality of $p(z)$---which are often broad, asymmetric, and/or
multimodal functions---against the delta functions represented by
spectroscopic redshifts.

\section{Measuring overconfidence}

\subsection{Conceptual explanation}\label{sec-HPDCI}

By its nature, $p(z)$ cannot be verified on a galaxy-by-galaxy basis,
just as a single coin toss cannot determine whether a coin is fair.  A
large sample, accordingly, does support $p(z)$ verification.  A sample
of 1000 galaxies, for example, {\it should} contain of order 10
galaxies whose spectroscopic redshift is in tension with the
photometric redshift at the 99\% level.  If too many galaxies exhibit
this much tension, the photometric redshifts collectively can be
deemed overconfident, as they predict the spectroscopic redshifts with
more precision than is supported by evidence.  Similarly, if too few
galaxies in the sample exhibit this much tension, the photometric
redshifts collectively can be deemed underconfident.

In practice, overconfidence is far more common than underconfidence,
when estimating almost anything.  This may be due to the nature of
error budgets: humans use judgment to identify the most salient
sources of uncertainty worthy of quantification, whereas
``subdominant'' sources of uncertainty have little effect when added
in quadrature and therefore do not merit quantification.  Sources of
uncertainty that initially do not seem salient may never be folded
into the budget even if their true contribution is substantial.  For
example, most photometry codes base their uncertainties on photon
noise and neglect sky modeling uncertainties.  Neglecting this source
of noise is justified in many cases, but the general pattern is to
underrepresent some sources of noise without any compensating
overrepresentation of other sources, so the final result is often
overconfident.

We can check for overconfidence by asking whether 50\% of galaxies
have their spectroscopic redshift within their 50\% credible
interval\footnote{Bayesian statisticians use this term when speaking
  of Bayesian posteriors, and reserve the term {\it confidence
    interval} for the likelihood.  This distinction does not often
  appear in the astronomy literature.}  (CI), 90\% have spectroscopic
redshift within their 90\% CI, etc.  Such checks do appear in the
literature \citep[e.g., ][]{Schmidt13,Dahlen2013zphot}, but are
usually implemented without a key feature that greatly assists with
the interpretation.  This key feature was evident already in the
pioneering work of \citet{Fernandez02}; here we explain it in more
detail, use it to implement a systematic confidence test, and show how
this test can lead to insight about the photometric redshift
algorithms themselves.

\begin{figure}
\centerline{\resizebox{3.5in}{!}{\includegraphics{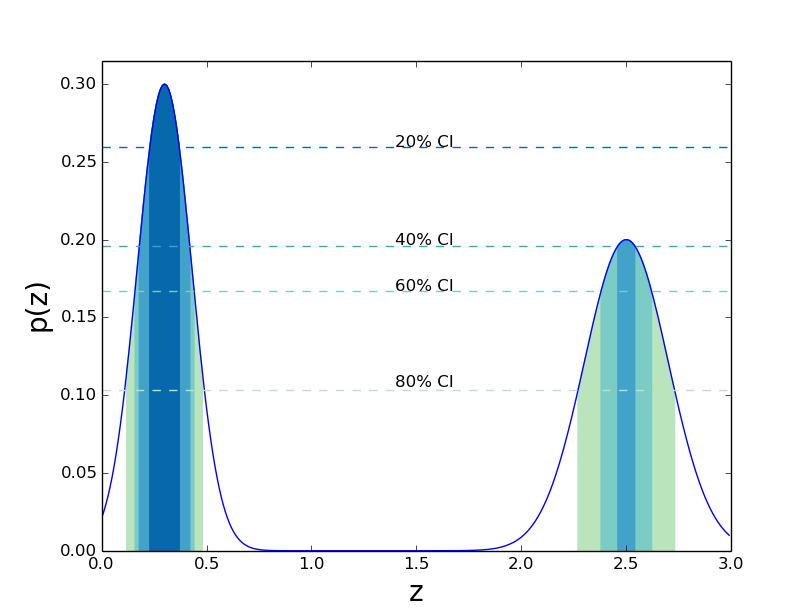}}}
\caption{Five illustrative highest probability density credible
  intervals. The darkest shade indicates the HPD 20\% CI, the next
  darkest shade (in combination with the first) indicates the HPD 40\%
  CI, and so on, with white comprising the final 20\%. The horizontal
  dashed lines indicate that the corresponding, possibly disjoint,
  redshift intervals have been identified starting from the peak by
  lowering a threshold and using the points where $p(z)$ crosses the
  threshold. \label{fig-pzexample}}
\end{figure}

For a confidence test, it is crucial that we choose the highest probability
density (HPD) CI for any given credibility level. To see why, consider
Figure~\ref{fig-pzexample}, which shows a hypothetical posterior
$p(z)$. We could define a 20\% CI by, say, starting at $z=0$ and
integrating the area under the posterior curve until we reach 20\% of
the total area under the curve, and indeed 20\% of galaxies should
have spectroscopic redshift within the 20\% CI as defined this
way. However, testing CIs defined this way would not test
overconfidence---the tendency for $p(z)$ to be too sharply peaked.  We
therefore define the 20\% CI by lowering a threshold from the peak
downward until the area under the parts of the curve intersected by
the threshold equals 20\% of the total area; this is the HPD 20\% CI.

The same process leads to the 40\%, 60\%, and 80\% CI in
Figure~\ref{fig-pzexample} covering multiple separated redshift
intervals.  This is the only way to maintain the highest probability
density and thereby probe for overconfidence.  To illustrate, imagine
this galaxy has $z_s=1.2$, a result that we should judge to be very
unlikely given this $p(z)$. This indeed falls outside the HPD 99\% CI,
but falls inside the 50\% CI if we define the CI by integrating only
in a single contiguous region around the highest peak.  The latter
definition of CI gives us a false sense that the photometric redshift
prediction was borne out.  An example with two equal peaks is
admittedly extreme, but the same principle is at work with unequal
peaks or even a single asymmetric peak.  Most confidence checks in the
literature to date have not used the HPD CI, but authors should begin
doing so; Section~\ref{sec-DLS} shows that results can differ
substantially when not using the HPD CI.  For the remainder of the
paper, references to CI should be understood as HPD CI unless
otherwise stated.


\subsection{Implementation}

We want to know how many galaxies in a data set have $z_s$ within
their 1\% CI, how many within their 2\% CI, etc. Computationally, this
implies a loop over credibility levels, with each iteration containing
a loop over galaxies to check whether each galaxy meets the criterion.
However, it is computationally more efficient to perform a one-time
calculation for each galaxy, to find the CI that just barely includes
the spectroscopic redshift.  Referring again to
Figure~\ref{fig-pzexample}, imagine that we have not yet calculated
any CI but we know the value of $z_s$. We simply draw a horizontal
line through $p(z_s)$ to identify the relevant redshift intervals and
compute the area under the curve in those intervals to find the
credible level that just includes $z_s$.  Recording that this
threshold credibility is, say, 32\% is a highly efficient way of
recording that this galaxy does not have $z_s$ within its 1\% CI, nor
its 2\% CI, nor its 3\% CI, etc, but does have $z_s$ within its 32\%
CI, and its 33\% CI, and its 34\% CI, etc.

The implentation is thus quite simple. We compute the threshold
credibility $c_i$ for the $i$th galaxy with:
\begin{equation}
c_i = \sum_{z \in p_i(z) \ge p_i(z_{s,i})} p_i(z)
\end{equation}
where $p_i(z)$ is the posterior for the $i$th galaxy, assumed to be
normalized.  The requirement that 1\% of galaxies have $z_s$ within
their 1\% CI, etc, then translates into a requirement that $c$ be
uniformly distributed from 0 to 1 (we drop the $i$ subscript when
referring to collective properties of the $c_i$).  We test for this by
computing the empirical cumulative distribution function $\hat{F}(c)$,
which should equal $c$.  Graphically, plotting $\hat{F}(c)$ resembles
a q-q plot in which $\hat{F}$ is expected to match $c$, i.e., fall on
a line through the origin with a slope of one. Overconfidence
corresponds to $\hat{F}(c)$ falling below this line (too few galaxies
have $z_s$ within a given CI).  The statistical significance of such a
departure can be measured with a Kolmogorov-Smirnov (KS) test.  Of
course, it is also possible for this test to reveal {\it
  under}confidence.  In either case, the method detects inaccurate
error budgets.

\section{Applying the test}

We tested the $p(z)$ estimated by two template-based photometric
redshift codes, BPZ \citep{BPZ} and EAZY \citep{Brammer08}.  We are
primarily interested in testing template methods because empirical
methods should yield calibrated $p(z)$ by design.  Because template
methods purport to yield redshifts beyond the magnitude limit at which
$p(z)$ can be directly constructed from the photometric and
spectroscopic data, they are the methods for which an independent test
of $p(z)$ is most desirable.  The test results will be data-dependent
and must be interpreted accordingly.  For example, any overconfidence
in the underlying photometry will contribute to overconfidence in
$p(z)$, and this contribution will be magnitude- and
redshift-dependent.  We therefore run each code on the same data, the
Hubble Deep Field North (HDFN) seven-band photometry with 127
spectroscopic redshifts \citep{FS99} that ships with EAZY and shipped
with earlier versions of BPZ.  We run each code with the default
templates and priors that are shipped with the code.

\subsection{BPZ}

\begin{figure}
\centerline{\resizebox{3.5in}{!}{\includegraphics{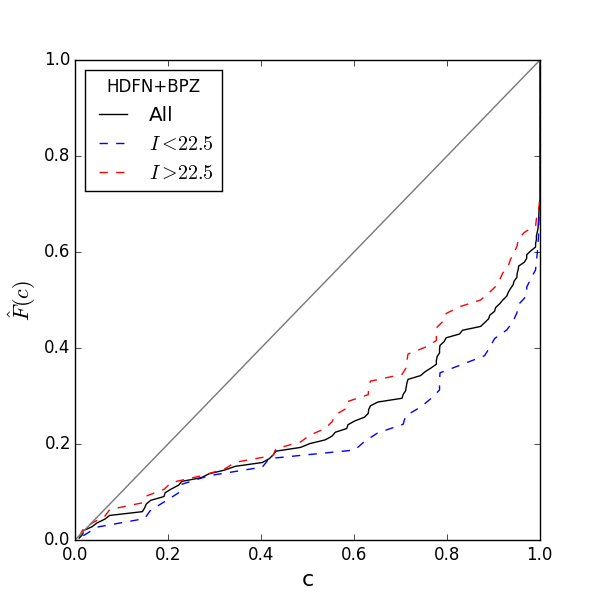}}}
\caption{The $\hat{F}(c)$ plot for HDFN data using the BPZ code shows
  substantial overconfidence overall (solid curve, $p< 10^{-15}$),
  with some magnitude dependence (dashed curves).
\label{fig-hdfn-bpz}}
\end{figure}

We used BPZ version 1.99.3 with default templates and priors and the
INTERP value set to 2 on the command line as recommended in the
documentation. The resulting $\hat{F}(c)$ plot
(Figure~\ref{fig-hdfn-bpz}, middle curve) shows substantial
overconfidence; the largest departure from the ideal distribution is
where only 46\% of galaxies have true redshift within their 89\% CI.
This is highly significant ($p<10^{-15}$) according to the KS test.

Next, we tested the magnitude dependence of this overconfidence by
breaking the sample into roughly equal bright ($I<22.5$) and faint
($I>22.5$) subsamples. The faint subsample falls on the upper curve in
Figure~\ref{fig-hdfn-bpz}, and the bright subsample falls on the lower
curve; in other words, there is more overconfidence in the brighter
galaxies.  A KS test indicates that the difference between the bright
and faint distributions is not significant ($p = 0.44$). However,
magnitude dependence will be a recurring issue so a few points should
be clarified now. First, magnitude-dependent overconfidence does not
automatically imply a problem in the photometric redshift algorithm,
because uncertainties in the underlying photometry are already
magnitude-dependent. Second, our method probes {\it how well the
  uncertainty has been assessed} rather than the uncertainties
themselves, so one should not assume that faint galaxies will be more
problematic.  Indeed, Figure~\ref{fig-hdfn-bpz} shows that faint
galaxies perform better.  Photometry provides a useful analogy: faint
galaxies have larger photometric uncertainties than bright galaxies,
but the faint-galaxy uncertainty budget is probably more accurate
because it is dominated by well-modeled photon noise rather than
poorly modeled uncertainties in background subtraction, calibration,
and color terms.  In fact, because photometry uncertainties propagate
into photometric redshift uncertainties, it is tempting to offer this
as an explanation for the overconfidence pattern seen in
Figure~\ref{fig-hdfn-bpz}.  However, the HDFN is a carefully
calibrated and well-tested catalog, and we develop a more compelling
explanation below.

Third, magnitude is correlated with redshift so disentangling the two
variables may be difficult. If the magnitude trend here is really a
redshift trend in disguise, splitting by redshift should reveal a
greater divergence between subsamples, and low-redshift (bright)
galaxies should have the most overconfidence.  In fact, splitting by
redshift shows the opposite in both respects, leading us to believe
that magnitude is indeed the explanatory variable.  An important
conceptual point here is that use of a Bayesian prior on redshift
prevents us from expecting all subsamples split by redshift to follow
the $\hat{F}(c)=c$ relation. In a Bayesian framework, priors can and
should degrade the performance of some subsets in order to improve
overall performance. As an everyday example, consider the batting
averages of baseball players one month into the season. With each
player having few at-bats, their averages vary widely, and applying a
prior on batting average greatly improves our estimate of their
``true'' batting averages.  But if, at the end of the season, we find
the players with the best ``true'' batting averages and look back at
our Bayesian estimate one month into the season, we will find that the
prior biased their averages low; this was unavoidable if we were to
improve the one-month estimates overall.  Similarly, high-redshift
subsamples must not be tested in isolation, and in this paper we do
not plot $\hat{F}(c)$ for subsamples split by redshift, even if we do
a redshift split to check whether a magnitude trend could be a
redshift trend in disguise.


We also tested BPZ using the test data it ships with, a catalog of 57
galaxies with spectroscopic redshift and seven-band photometry from
the Hubble Ultra Deep Field (HUDF) catalog produced by \citet{Coe06}.
We found trends similar to those illustrated here for the HDFN data,
suggesting that overconfidence is a general feature of the $p(z)$
output by BPZ.  A plausible mechanism for this is that BPZ, like most
template codes, propagates uncertainty from the photometry only, and
not from the templates. This would also explain why overconfidence is
greater for bright galaxies; their smaller photometric uncertainties
imply that template uncertainties are a larger share of the
uncertainty budget.  Further supporting this picture, we found that
limiting the template set by turning off interpolation between
templates (setting the INTERP parameter to 0) exacerbated the
overconfidence.  Increasing the INTERP parameter beyond 2 had little
effect, presumably because SEDs vary in ways that cannot be captured
with interpolation between the default templates.

The issue of template uncertainty was recognized by
\citet{Fernandez02}, who explored an empirical fix of convolving
$p_i(z)$ with a Gaussian smoothing kernel.  Although they cautioned
that more sophisticated noise modeling would be required as data sets
expanded in terms of both redshift and raw numbers, this approach
performed well when they applied it to HDFN data.  They assumed that
the kernel width should scale as $(1+z)$ and then optimized the
prefactor by maximizing $p_i(z_s)$ for bright galaxies.  This yielded
a kernel with $\sigma=0.065(1+z)$; this kernel is ``optimal'' in the
sense that it mimics the effect of galaxy SEDs varying from the
template set used in their analysis, at the wavelengths used in their
analysis, better than other kernels in its family.  Smoothing with
this kernel broadens $p_i(z)$ for each galaxy, but more so for bright
galaxies because faint galaxies already have broad $p_i(z)$ due to
their photometric uncertainties.  \citet{Fernandez02} tested the
performance of this procedure with a version of the $\hat{F}(c)$ test,
verifying (in our notation) $\hat{F}(0.683)$, $\hat{F}(0.954)$, and
$\hat{F}(0.997)$.  

BPZ does have two parameters that nearly serve this function.  The
CONVOLVE\_P parameter, if set, smooths $p_i(z)$ with a Gaussian of
fixed width $\sigma=0.03$.  According to comments in the code the
purpose of this feature is to combine multiple close peaks; for our
purposes we can consider it as injecting a bit of template noise, but
Figure~\ref{fig-hdfn-bpz} already includes this bit of smoothing
because CONVOLVE\_P is turned on by default. Therefore $\sigma=0.03$
is too little smoothing to prevent overconfidence, at least for the
data sets presented here.  The other potentially relevant BPZ
parameter is MIN\_RMS, which according to the comments represents
``intrinsic photo-z rms'' (presumably due to the true SEDs of galaxies
varying from the templates). In version 1.99.3 MIN\_RMS is set to 0.05
(0.067 if the older ``CWWSB'' template set is used) but it does {\it
  not} affect the $p(z)$ written out to disk.  It is used only to
determine a few quantities derived from $p(z)$, such as upper and
lower redshift limits and the fraction of the area under $p_i(z)$ that
is near the highest peak.

\begin{figure}
\centerline{\resizebox{3.5in}{!}{\includegraphics{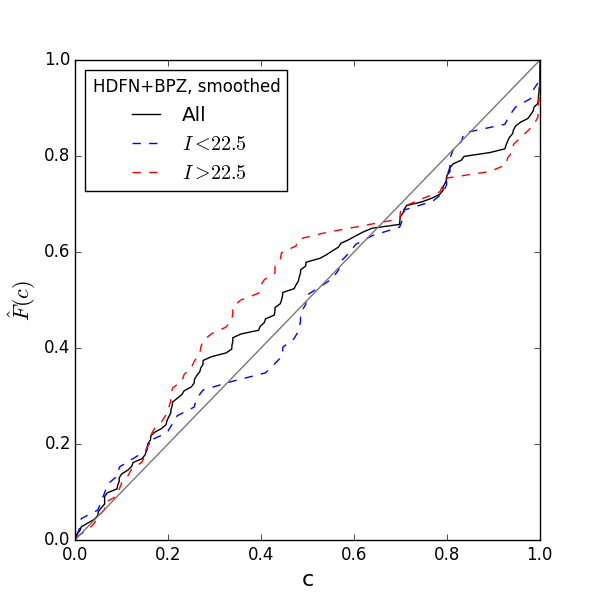}}}
\caption{The $\hat{F}(c)$ plot for HDFN data using the BPZ code plus
  post-hoc $p_i(z)$ smoothing with a Gaussian kernel with
  $\sigma=0.11$.  All the resulting distributions are at least
  marginally consistent with uniformity, suggesting that 
 smoothing $p_i(z)$ is a reasonable substitute for modeling template noise in BPZ.  
 \label{fig-bpz2}}
\end{figure}

We therefore use a post-processing step to smooth the $p_i(z)$
produced by BPZ in order to test the efficacy of the approach
suggested by \citet{Fernandez02}.  In the absence of strong evidence
that a redshift-dependent kernel is necessary, we tested
redshift-independent kernels of various widths.  We found that a
Gaussian kernel with $\sigma=0.11$ was optimal in the sense of
balancing overconfidence with underconfidence, as seen in
Figure~\ref{fig-bpz2}.  The resulting $\hat{F}(c)$ is marginally
consistent with uniformity for the overall sample and the faint
subsample ($p=0.07$ and 0.04 respectively), and entirely consistent
for the bright subsample ($p=0.89$).  Given the median redshift (0.75)
of the HUDF spectroscopic sample, this agrees well with the
$\sigma=0.065(1+z)$ derived by \citet{Fernandez02} for the HDFN
sample.  We performed the same tests on the $p_i(z)$ output by BPZ for
the HUDF sample and found the same result.  Thus, smoothing $p_i(z)$
with a standard kernel may be an adequate substitute for modeling
template noise in many situations.

\subsection{EAZY}\label{subsec-EAZY}


The EAZY code models template uncertainties as a function of
rest-frame wavelength.  This approach, called the template error
function, has several virtues.  First, there is strong physical
motivation for assuming that template variance is a function of
rest-frame wavelength \citep{HH10}.  Second, any effect that depends
on rest-frame wavelength propagates into $p_i(z)$ in a filter- and
redshift-dependent way that cannot be fully mimicked by simply
broadening $p_i(z)$.  For example, consider a bimodal $p_i(z)$ with
two similar well-separated peaks.  The smoothing approach will blindly
broaden both peaks, but the template error function may effectively
broaden one peak much more than the other, according to the relevant
rest-frame template uncertainties.  \citet{Brammer08} calibrate their
template error function using a bright galaxy subsample with known
spectroscopic redshifts, and provide the tools to recalibrate the
error function if desired.

\begin{figure}
\centerline{\resizebox{3.5in}{!}{\includegraphics{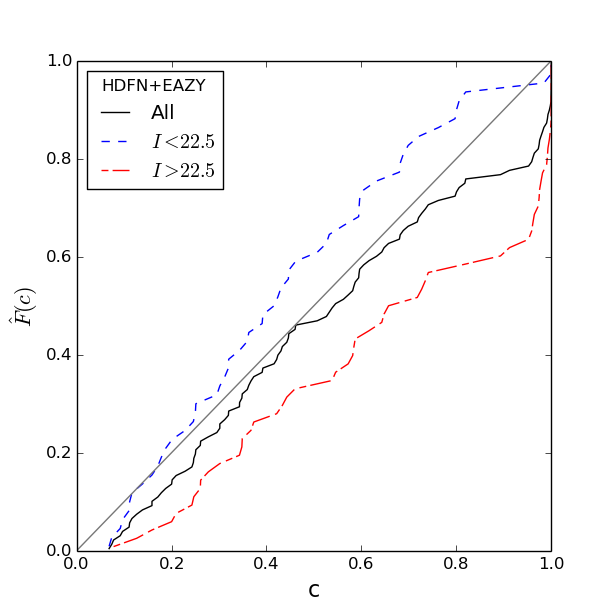}}}
\caption{When applied to the HDFN (solid curve), EAZY produces $p(z)$
  that are only slightly overconfident (but this is statistically
  significant: $p=0.0029$). The magnitude trend is inconsistent with
  unmodeled template uncertainty, and could be a redshift
  effect. \label{fig-eazy}}
\end{figure}

EAZY ships with the HDFN data, and Figure~\ref{fig-eazy} shows the
resulting $\hat{F}(c)$.  For the entire sample (solid curve), there is
only a modest amount of overconfidence, with a maximum deviation of
0.167 from the identity relation (78\% of galaxies are within their
94\% CI).  According to the KS test, this meets standard criteria for
statistical significance, but it is not overwhelming ($p=0.0029$).
Splitting into roughly equal subsamples by magnitude (dashed lines)
reveals a substantial magnitude dependence, with overconfidence on the
faint subsample and a bit of underconfidence on the bright subsample.
To check whether this magnitude dependence could really be a redshift
dependence, we split by redshift and find an even larger difference,
in the sense of even more underconfidence for low-redshift galaxies
than for bright galaxies.  This implies that redshift could be the
driving variable here; and as explained above, redshift variations in
these tests could simply reflect the workings of the Bayesian prior on
redshift.  Thus, the magnitude dependence may be a feature rather than
a flaw.

The relatively good performance of the overall sample and the lack of
overconfidence in the bright subsample indicate that the template
error function is generally serving its purpose in EAZY.  Some
overconfidence remains in the overall sample, but the good performance
of the bright subsample suggests that this is not due to
overconfidence in the templates. The remaining overconfidence may stem
from other aspects of the algorithm or from the photometry; for
example, unmodeled uncertainties in deblending or local background
variations may affect faint galaxies more than bright galaxies and
thereby fit this pattern. Investigating this hypothesis is beyond the
scope of this paper, but we offer some recommendations in
Section~\ref{sec-discussion}.

\section{Application to Deep Lens Survey}\label{sec-DLS}

\begin{figure}
\centerline{\resizebox{3.5in}{!}{\includegraphics{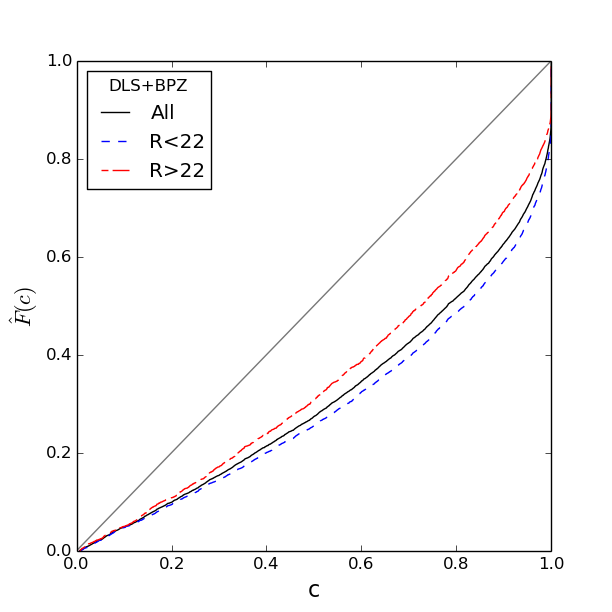}}}
\caption{The $\hat{F}(I^p)$ plot for BPZ applied to DLS photometry of
  the PRIMUS spectroscopic sample shows overconfidence
  ($p<10^{-300}$), preferentially in the brighter galaxies.  This
  suggests the need to model template uncertainties.\label{fig-dls}}
\end{figure}

The Deep Lens Survey \citep[DLS; ][]{DLS02,Wittman06} is a
ground-based 20 deg$^2$ {\it BVRz} survey that provides a counterpoint
to the HDFN and HUDF samples in terms of photometric uncertainties
(larger from the ground) and filter set (DLS uses a minimal filter set
to maximize the area covered).  \citet[hereafter ST13]{Schmidt13}
describe the DLS photometric redshifts and verify them using
$\sim 10^4$ spectroscopic redshifts in a 1 deg$^2$ overlap region with
the Prism Multi-object Survey \citep[PRIMUS,][]{Primus}.  ST13 tested
the fraction of galaxies with spectroscopic redshift within six
different CI and did not find systematic under- or overconfidence.
However, our $\hat{F}(c)$ test on these data (Figure~\ref{fig-dls})
reveals substantial overconfidence.  The maximum departure from the
desired line is nearly 0.30---only 53\% of the galaxies are within
their 82\% CI---which a KS test deems overwhelmingly significant
($p<10^{-300}$).  Because of the large sample size---8719 galaxies
after applying all the cuts applied by ST13---all departures visible
in the DLS plots in this paper are significant.

The difference between the ST13 results and ours lies in the
definition of CI. ST13 integrated around the highest peak and
integrated symmetrically (in terms of area under $p_i(z)$) around that
peak.  Therefore, their CI are not HPD CI and should not be used to
probe for overconfidence, as explained in Section~\ref{sec-HPDCI}.

\begin{figure}
\centerline{\resizebox{3.5in}{!}{\includegraphics{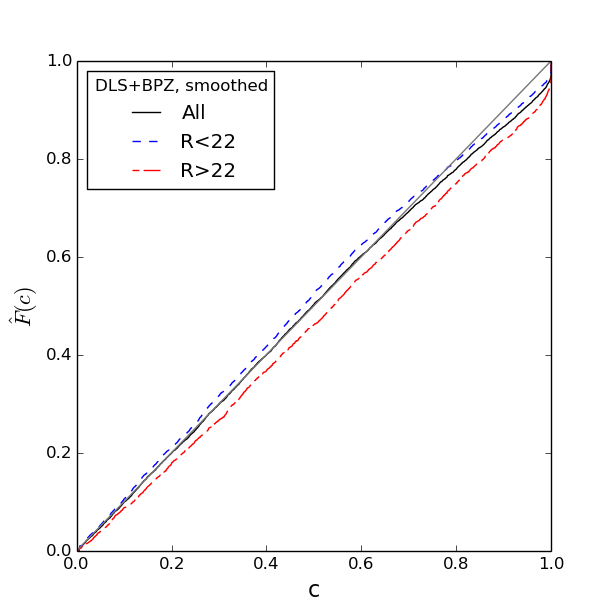}}}
\caption{The $\hat{F}(c)$ plot for BPZ+DLS with $p_i(z)$ smoothing
  using $\sigma=0.055$.  The smoothing provides a good approximation
  to the desired behavior, with a small magnitude trend.  All
  departures from the line are statistically significant.
\label{fig-dls-smooth}}
\end{figure}

Next, we probe for trends by splitting the data into subsamples.  We
found no visible difference in subsamples split by spectroscopic
redshift and only small differences in subsamples split by galaxy
spectral type (as determined by BPZ).  We did find a modest trend with
magnitude, with brighter galaxies showing more overconfidence
(Figure~\ref{fig-dls}, dashed curves), the same trend exhibited by BPZ
on the HDFN and HUDF data.  The sign of the magnitude trend, along
with the lack of a redshift trend, again points to template noise.  We
therefore tried the $p_i(z)$ smoothing approach with a series of
kernel widths, and in Figure~\ref{fig-dls-smooth} we plot the results
using $\sigma=0.055$ to illustrate the difficulty of defining an
``optimal'' kernel. The $\sigma=0.055$ kernel shown here provides a
quick way to remove most of the overconfidence, but leaves a
discrepancy in the tails: 92.7\% of galaxies are within their 97.2\%
CI ($p\approx 10^{-15}$).  A broader kernel would reduce the
discrepancy in the tails, but would also introduce {\it
  under}confidence elsewhere in the plot. Choosing the kernel by the
sole criterion of minimizing the deviation between $\hat{F}(c)$ and
$c$ may not be wise here because the bright subsample is already
generally underconfident after smoothing, {\it except} in the
tails. This suggests that template noise has been ``modeled'' about as
well as it can be with a Gaussian smoothing kernel, and that an
optimal kernel would include heavier tails.  Note also the remaining
overconfidence in the {\it faint} galaxies: this suggests that some
photometric uncertainties remain unmodeled.

\begin{figure}
\centerline{\resizebox{3.5in}{!}{\includegraphics{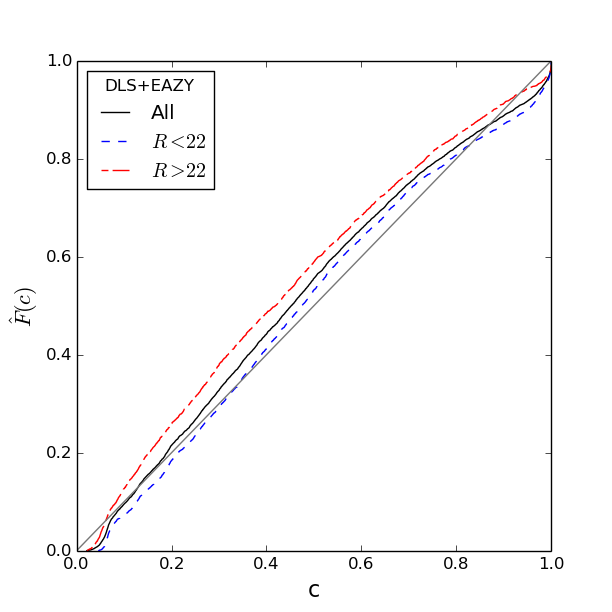}}}
\caption{The $\hat{F}(c)$ plot for EAZY applied to DLS photometry of
  the PRIMUS spectroscopic sample shows a relatively small (but
  significant) amount of overconfidence (93.2\% of galaxies are within
  their 96.8\% CI, $p\approx 10^{-10}$).  As with the smoothed BPZ
  results, the remaining overconfidence is primarily in the tails,
  suggesting that template uncertainty is not the cause of the
  overconfidence.\label{fig-dlseazy}}
\end{figure}

We also ran EAZY on the DLS data.  The $\hat{F}(c)$ curve
(Figure~\ref{fig-dlseazy}) is much better than for BPZ on the same
data, but still departs significantly from the desired distribution in
places (93.2\% of galaxies are within their 96.8\% CI,
$p=\approx 10^{-10}$). There is also some {\it under}confidence where
the empirical curves pass above the diagonal line in
Figure~\ref{fig-dlseazy}.  

Although the EAZY and smoothed BPZ results have many dissimilarities
(e.g., opposite magnitude trends), they both have a kink in the curve
at $c\approx 0.8$: spectroscopic redshifts too frequently land very
far from the $p(z)$ peak.  This suggests non-Gaussian wings in the
photometric uncertainties (from, e.g., deblending), the template
uncertainties, or both.  As discussed at the end of
Section~\ref{subsec-EAZY}, these two sources of uncertainty can be
decoupled by modeling the photometric uncertainties---including heavy
tails---via simulations and repeat visits.  Photometric redshift
outliers can be substantially reduced simply by folding this
heavy-tailed photometric model into a standard code such as BPZ
\citep{Wittman07}.  Outliers remaining after this process are likely
due to non-Gaussian template uncertainties.

Finally, we note that the DLS results presented so far could reflect
differences between EAZY and BPZ other than the template error
function:
\begin{itemize}
\item EAZY fits for a linear combination of templates with nonnegative
  coefficients, while BPZ considers templates separately (but
  interpolates between successive templates).
\item BPZ has a type-dependent prior while EAZY does not. We ran EAZY
  on the DLS data using a prior as similar as possible to the ST13 BPZ
  prior, but due to this restriction the priors cannot be the same.
\item We also used different templates. For EAZY we used the EAZY 1.0
  templates 1--6, whereas the BPZ $p_i(z)$ we downloaded from the DLS
  data release website\footnote{\url{http://dls.physics.ucdavis.edu}}
  resulted from the more highly tuned ST13 templates.
\end{itemize}
We therefore tested whether the template error function is the primary
factor responsible for reduced overconfidence in EAZY by running EAZY
on the DLS data with the template error function turned off.  We
obtained an $\hat{F}(c)$ distribution remarkably similar to the BPZ
distribution shown in Figure~\ref{fig-dls}, indicating that the
template error function is indeed the primary cause of reduced
overconfidence.  Even with the template error function on,
overconfidence rose when we restricted or eliminated EAZY's linear
combination feature.  This suggests that two conditions must be met
for appropriate confidence.  First, the available templates must be
able to match the gross features of the observations, either by
combining many generic templates as EAZY does by default, or by
tweaking a smaller number of templates as ST13 did before applying BPZ
to the DLS.  Second, spectral energy variations on finer wavelength
scales can be modeled via the template error function.

\section{Summary and discussion}\label{sec-discussion}

Applications of photometric redshifts increasingly use each galaxy's
probability density function $p_i(z)$ rather than a single point
estimate, and rightly so. However, tests of photometric redshift
accuracy generally compare only the highest $p_i(z)$ peak to the
spectroscopic redshift.  This can lead to the false conclusion that a
galaxy is a ``catastrophic outlier'' when in fact it is entirely
predictable that a spectroscopic redshift sometimes falls on the
second-highest peak, or in even lower-probability regions.  We have
defined a way to test all parts of $p_i(z)$ using the empirical
cumulative distribution function $\hat{F}$.  This test determines the
collective consistency of the $p(z)$ with the spectroscopic redshifts
in way that probes specifically for the known failure mode of
overconfidence.  The test complements rather than replaces traditional
tests because the latter are still necessary for measuring differences
in terms of redshift.

We find that the $p(z)$ produced by BPZ (and presumably most other
template methods) suffer from substantial overconfidence (e.g. only
32\% of galaxies have true redshift within their 92\% CI) because they
do not account for variation in galaxy SEDs, so-called template
noise. One code that does model template noise, EAZY, produced $p(z)$
with substantially less overconfidence, on each of two data sets that
differ widely in terms of filter set and depth.  Multiple independent
arguments suggest that the improved performance is due to template
uncertainty modeling rather than other differences between the codes.
First, the most marked difference between BPZ and EAZY is with {\it
  bright} galaxies, for which template noise is a larger fraction of
the uncertainty budget and for which priors should be relatively
unimportant.  Second, smoothing the BPZ $p_i(z)$---a crude model of
the effect of template uncertainty---greatly reduces the BPZ
overconfidence. Third, we turned the template uncertainty modeling off
in EAZY and found overconfidence similar to BPZ.

The practical impact of this overconfidence is not immediately
apparent from statements about the percentage of galaxies within a
given CI.  On an individual galaxy level, the practical impact is that
the true redshift is not as well constrained as the $p_i(z)$ would
indicate, and we can quantify this by specifying the amount of
$p_i(z)$ smoothing required to eliminate the overconfidence (to the
extent possible with smoothing rather than with more sophisticated
modeling). We found that smoothing the BPZ $p_i(z)$ with $\sigma=0.09$
is adequate for the HUDF data set, and $\sigma\approx 0.06$ is
adequate for the DLS; both numbers are consistent with the suggestion
of $\sigma=0.065(1+z)$ by \citet{Fernandez02}. The ``optimal'' kernel
width may depend on filter set and other data details, as it reflects
how much the SEDs vary from the templates at the rest wavelengths most
heavily probed by the data, but $\sigma=0.065(1+z)$ seems to work with
a variety of deep optical surveys.  A fixed kernel does have the
flexibility to work with varying numbers of filters, because it has
appropriately less impact on the already-broad $p_i(z)$ produced by
surveys with few filters.

For sets of galaxies such as a $z_{p}$ bin used in cosmic shear, the
practical effect of overconfidence is that the true redshift
distribution of the set is likely to be broader than the summed
$p_i(z)$.  For a hypothetical distribution of galaxies centered at
$z=1$ and with $\sigma_z=0.2$, to first order the effect of smoothing
with a $\sigma=0.065(1+z)$ kernel will be to broaden the bin to
$\sigma_z=0.24$. For weak lensing tomography with a next-generation
imaging survey like LSST \citep{LSST}, \citet{Ma2006} report that the
width of the redshift bins must be known to better than 0.01 to avoid
substantial degradation of dark energy parameter constraints (where
substantial degradation is defined as parameter constraints 1.5 times
looser than in the case with perfect knowledge of the bin width).
Accurate modeling of template uncertainty is therefore likely to be
important in achieving the full potential of such surveys.  In doing
so, we must avoid the traditional anti-overconfidence tactic of
multiplying the error bars by some factor in order to adopt a
``conservative'' estimate.  Broadening $p(z)$ too much
(underconfidence) results in overestimating the width of a redshift
bin, which is equally harmful to much of the downstream science.

Although we have focused on smoothing the $p_i(z)$ as a convenient
fix, template variance at a given rest wavelength propagates into
different observed filters at different redshifts.  Physical modeling
of this process is therefore better than smoothing in principle.  In
either case, correcting this source of overconfidence helps expose
other issues with the overall uncertainty budget.  This is best
illustrated by the comparison between Figures~\ref{fig-dls} and
\ref{fig-dls-smooth}: an excess of spectroscopic redshifts very far
from the $p(z)$ peaks is clearly visible in the upper right corner of
Figure~\ref{fig-dls-smooth} but is masked by the much larger
overconfidence trend in Figure~\ref{fig-dls-smooth}.  This pattern of
outliers appears across a variety of codes and data sets, and points to
the need to model heavy tails in probability distributions, including
that of the underlying photometry.  For large data sets and surveys we
recommend decoupling photometric uncertainty from other issues by
conducting targeted simulations and repeat observations to carefully
calibrate the photometric uncertainty model.  Better photometric
uncertainty modeling will yield better photometric redshifts
\citep{Wittman07}, and in turn will enable photometric redshift
confidence calibration to focus on physical modeling components such
as templates.  Because the photometric noise contribution is strongly
magnitude-dependent, both types of modeling will be necessary to
understand a survey over its full magnitude range.

The probability tails are potentially important for downstream
science, because a small leakage of high-redshift galaxies into a
low-redshift bin could add substantially to the naturally low lensing
signal in that bin, while a small leakage in the other direction can
substantially change the inferred luminosity function at high
redshifts.  Tracking these details requires tools other than the
overconfidence test; for example, the leakage can be mapped with a
$z_s$ vs. $z_p$ plot in which $z_p$ is rendered as a cloud
corresponding to $p(z)$.  In the end, the true redshift distribution
of a photometric redshift bin may best be constrained by methods that
are independent of any photometric redshift algorithm \citep[e.g. the
cross-correlation method,][]{Newman08}.

The overall uncertainty budget is strongly magnitude-dependent, so
tests performed with bright spectroscopic samples should be
interpreted carefully.  Obtaining a truly representative spectroscopic
subsample is difficult; for example, the PRIMUS spectroscopy in the
DLS field has a 50\% redshift success rate at $R\approx 21.5$
\citep{Cool13} while the DLS photometry goes much deeper than
that. Our conclusions regarding template uncertainty are robust,
however, because the spectroscopic sample is most complete for bright
galaxies, where template variance is the largest fraction of the
uncertainty budget.

\section*{Acknowledgements}

We thank Sam Schmidt for valuable discussions and help with running
EAZY on the DLS data.  We also thank Paul Baines and Karen Ng for
useful discussions, and the anonymous referee for helpful feedback.
The DLS was made possible by support from Lucent Technologies and NSF
grants AST 04-41072 and AST 01-34753.

\bibliographystyle{mnras}
\bibliography{ms}

\end{document}